\documentclass[12pt,letterpaper]{article}
\usepackage[affil-it]{authblk}
\usepackage{amsmath, amsfonts, amsthm, amssymb}
\usepackage[usenames,dvipsnames]{color}
\usepackage{comment}
\usepackage{ifpdf}
\usepackage{setspace}
\usepackage[utf8]{inputenc}
\usepackage[english]{babel}
\usepackage{graphicx}  
\usepackage{breqn}
\usepackage{mathrsfs}
\usepackage[left=.8in,right=.8in,top=.8in,bottom=.8in]{geometry}     
\usepackage{slashed}
\usepackage{mathtools}
\usepackage{amsfonts}
\usepackage{amssymb}
\usepackage{epsfig}
\usepackage{rotating}
\usepackage{url}
\usepackage{times}
\usepackage{color}
\usepackage{bm}
\usepackage{xcolor,colortbl}
\usepackage{hyperref}
\usepackage[alsoload=hep]{siunitx}
\usepackage{enumitem}
\usepackage{fancyhdr}
\usepackage{anyfontsize}
\usepackage{wasysym}
\usepackage{empheq}

\newcommand{\nn}{\nonumber}

\newcommand{\be}{\begin{equation}}
\newcommand{\ee}{\end{equation}}
\newcommand{\bea}{\begin{eqnarray}}
\newcommand{\eea}{\end{eqnarray}}

\newcommand{\sign}{\textrm{sign}}

\begin{document}


\title{A generalized Hartle-Hawking wave function}

\author{Stephon Alexander and Gabriel Herczeg}
\affil{Department of Physics, Brown University, Providence, RI, 02906, USA}

\author{Jo\~{a}o Magueijo}
\affil{Theoretical Physics Group, The Blackett Laboratory, Imperial College, Prince Consort Rd., London, SW7 2BZ, United Kingdom}
\date{\today}

\maketitle

\begin{abstract}
The Hartle-Hawking wave function is known to be the Fourier dual of the Chern-Simons or Kodama state reduced to mini-superspace, using an integration contour covering the whole real line. But since the Chern-Simons state is a solution of the Hamiltonian constraint (with a given ordering), its Fourier dual should provide a solution (i.e. beyond mini-superspace) of the Wheeler DeWitt equation representing the Hamiltonian constraint in the metric representation. We write down a formal expression for such a wave function, to be seen as the generalization beyond mini-superspace of the Hartle-Hawking wave function. Its explicit evaluation (or simplification) depends only on the symmetries of the problem, and we illustrate the procedure with anisotropic Bianchi models and with the Kantowski-Sachs model. A significant difference of this approach is that we may leave the torsion inside the wave functions when we set up the ansatz for the connection, rather than setting it to zero before quantization. This allows for quantum fluctuations in the torsion, with far reaching consequences. 
\end{abstract}

\section{Introduction}

Quantum gravity has historically displayed a schism between formalisms giving primacy to the metric,
and those that place the connection at the centre. While classically there is little difference between these contrasting approaches (at least if torsion vanishes),
the fact that quantum mechanics probes the off-shell phase space suggests that inequivalent quantum theories may follow from them. 
The attitude in this paper is that this should not impede communication
and cross-pollination between the two frameworks. Specifically, we will show how the Chern-Simons (Kodama) 
solution~\cite{jackiw,witten,kodama,lee1,lee2} in 
Ashtekar's connection-driven formulation may be used to generalize the Hartle-Hawking metric-based wave function of the Universe~\cite{H-H}.
By ``generalizing,'' we mean going beyond its mini-superspace origins (see~\cite{CSHHV}):
 we will formally obtain counterparts to the Hartle-Hawking state 
in any non-perturbative situation,
including anisotropic and inhomogeneous cosmological models, black holes and exact gravitational waves.

The Ashtekar formalism is a connection based approach to Quantum Gravity  where 
the Einstein-Cartan spin connection is re-encoded in an $SU(2)$ complex self-dual connection, leading to an elegant
formulation of Einstein gravity \cite{PhysRevLett.57.2244}. One motivation of the approach is to establish bridges with non-perturbative quantization methods used
in non-Abelian gauge theories (leading to loop quantum gravity). However, it is also possible to discuss quantization directly in terms of a representation 
diagonalizing the connection. It is in this context that the Chern-Simons state appears as a non-perturbative solution to the theory. 
Although this state has attracted significant criticism~\cite{witten,Randono0,Randono1,Randono2} its reassessment in a formalism that keeps the action and constraints manifestly real can resolve these issues~\cite{MSS,realK}. 

Curiously, it was only recently noted~\cite{CSHHV} 
that when the Chern-Simons state is reduced to mini-superspace it becomes the Fourier dual of the Hartle-Hawking
wave function of the Universe. 
Indeed, in \cite{Randono Mesoscopic} the Fourier transform of the mini-superspace Chern-Simons state was carried out explicitly
(with very interesting phenomenological implications), but the result was not recognized as the Hartle-Hawking state.
This result should have been obvious from the outset, however, 
given that the connection and the (densitized inverse) metric are complementary quantum variables, and the Chern-Simons state and the 
Hartle-Hawking wave function 
solve the {\it same} constraint equation written in terms complementary variables. 

The starting point of this paper is the remark that the Chern-Simons solution is not confined to mini-superspace. Furthermore, the Fourier transform between 
dual variables is also completely general.
It is therefore possible to define a metric
representation wave function dual to the Chern-Simons state in a general setting, 
and view this as the full non-perturbative, non-symmetry-reduced generalization of the Hartle-Hawking 
wave function. Such is the purpose of this paper. 

Issues will inevitably appear: foremost, the question of which contour to take in the integration. Here we take the minimalistic view that all variables to emerge from a canonical framework based on a real action should be real and cover the whole real line. A straightforward application of the Fourier theorem is then possible. But it is possible that a more detailed analysis permits forays into the complex domain, or that truncations of the real line for some variables are possible. This is the reason why we will confine ourselves to the dual of the Hartle-Hawking state and not the Vilenkin wave function~\cite{vil0,Vilenkin}. The matter can already be understood in mini-superspace~\cite{CSHHV}.  

Another issue concerns how to deal with torsion. Torsion is zero by construction in the metric approach, but not in any approach based on the 
connection. Of course we may set torsion to zero by hand in the classical Einstein-Cartan theory (in the absence of spinors), and then quantize, but this is not the 
only approach. Indeed it may be desirable to leave the second class constraints forcing the torsion to zero unsolved until a later stage in the quantum
analysis. This is the view advocated in~\cite{QuantTorsion} and we shall follow it in this paper. It will also serve as the basis of a future analysis~\cite{QuantFlat} of the 
flatness, anisotropy and singularity problems, as we outline in the conclusions to this paper.

\section{The basic idea}
For the sake of simplicity, our starting point is a formalism where the Chern-Simons state emerges from a manifestly real action, Hamiltonian, and phase space~\cite{realK}. 
This is possible by differentiating between the Immirzi parameter used in the definition of the connection (which we take to be $\gamma_1=i$)
and the one appearing in the pre-factor of the Holst term (which we set to infinity, i.e. no Holst term added). The Chern-Simons wave function and 
the commutator between complementary variables (and consequently the suggested transform between duals) then 
have the required properties to fall under the remit of the Fourier theorem. 

More concretely, given that the Einstein-Cartan action can be written as:
\be\label{EC}
S_{EC}=\kappa \int dt \, dx^3 \, \left[2\Im \dot A^i_a E_i^a-(NH + N^a H_a + N_i G^i) \right]
\ee
(where  $\kappa = 1/(16\pi G)$)
we have the Poisson bracket relations:
\be\label{PBnonpert}\{\Im A^i_a(\vec x),E^b_j(\vec y)\}=\frac{1}{2\kappa}
\delta^b_a\delta^i_j\delta(\vec x-\vec y)\;  \ee 
implying upon quantization:
\be\label{Comm}\left[\Im A^i_a(\vec x),E^b_j(\vec y)\right]=i l_P^2
\delta^b_a\delta^i_j\delta(\vec x-\vec y)\ee
(where $l_P= \sqrt{8\pi G\hbar}$ is the reduced Planck length).
Hence, in a representation diagonalizing the connection (and where $\Re A^i_a$ is seen as a parameter), we have $\psi_A(A^i_a(\vec x))=\langle A^i_a(\vec x)|\psi\rangle$ as well as $\hat A\psi_A(A)= A\psi_A(A)$ and 
\be
\hat E^a_i(\vec x)=-i l_P^2\frac{\delta}{\delta \Im A^i_a(\vec x) }.
\ee
In the complementary representation we have instead
$\psi_E(E^a_i(\vec x))=\langle E^a_i (\vec x)|\psi\rangle$,  $\hat E\psi_E(E)= E\psi_E(E)$, and:
\be
\hat \Im \hat A^i_a(\vec x)=i l_P^2\frac{\delta}{\delta E^a_i(\vec x) }.
\ee
For any delta-normalizable wave function, it follows that:
\be\label{FT}
\psi_E =\prod_{\vec x,a,i} \int \frac{d [\Im (A^i_a(\vec x))]}{\sqrt{2\pi l_P^2}} e^{-\frac{i}{l_P^2} E^a_i(\vec x) \Im A^i_a (\vec x)}\psi_A,
\ee
where we assume the integrals are over the real line. It is the last assumption (on the range of integration) that will pick out the Hartle-Hawking
boundary conditions in mini-superspace, rather than Vilenkin's~\cite{CSHHV}, as we already alluded. We make that assumption to avoid issues over the reality 
conditions (we are working with a theory which is manifestly real from the outset), but also to make sure the Fourier theorem applies outright.

But we know the {\it general} solution to the constraints contained in (\ref{EC}) in the $A$ representation for this explicitly real theory. It is the modification of the Chern-Simons state described in~\cite{realK}. The Hamiltonian constraint equation (i.e. the Wheeler-DeWitt equation) in the connection representation for an explicitly real theory (with a standard ordering) reads:
\be
\left ( \Re B^{kc} -\frac{i l_P^2 \Lambda}{3} \frac{\delta}{\delta \Im A^k_c(\vec x) } \right)\psi=0 \label{HamConst}
\ee
where $B^{kc} = \frac{1}{2}\epsilon^{abc}F^k_{ab}$, and $F^k_{ab}$ is the field strength tensor
\be
F^i_{ab} = \partial_a A^i_b - \partial_b A^i _a + \epsilon^{ijk}A^j_a A^k_b.
\ee

 A solution of \eqref{HamConst} is given by
\bea\label{kodrealth}
\psi_{CS}(A)&=&{\cal N}\exp {\left(\frac{3i}{ l_P^2\Lambda}\Im Y_{CS}\right)}, \label{CS state}
\eea
with
\bea
Y_{CS}&=&\int  {\rm Tr} \left(A dA +\frac{2}{3} A A  A\right)\nn\\
&=&-\frac{1}{2} \int  A^i dA^i +\frac{1}{3}\epsilon_{ijk} A^i A^j A^k.
\eea
To see that \eqref{CS state} solves \eqref{HamConst}, it is convenient to decompose ${A^i = \alpha^i + i\beta^i}$, with $\alpha^i$ and $\beta^i$ real. It is then straightforward to compute
\be
\Im Y_{CS} = -\int d\alpha^i \beta^i + \frac{1}{2}\epsilon_{ijk} \left(\beta^i\alpha^j\alpha^k - \frac{1}{3}\beta^i\beta^j\beta^k\right) 
\ee
where we have integrated by parts and discarded a boundary term to obtain the last expression, whence
\bea
\frac{\delta (\Im Y_{CS})}{\delta \beta^k_c} &=&  -\frac{1}{2}\epsilon^{abc}\left[\partial_a\alpha^k_b - \partial_b\alpha^k_a + \epsilon^{ijk}(\alpha^i_a\alpha^j_b - \beta^i_a\beta^j_b)\right] \nn \\
 &=& -\Re B^{kc},
\eea
from which it can be easily established that \eqref{CS state} solves \eqref{HamConst}.

By inserting (\ref{kodrealth}) into (\ref{FT}), we obtain the generalization for metrics with any symmetry (or indeed without any symmetry at all)
of the Hartle-Hawking wave function.

\section{Reduction to mini-superspace and the role of quantum torsion}
It is straightforward to see that when our proposal is applied to mini-superspace it reduces to~\cite{CSHHV}, where it is shown 
that the Chern-Simons state is the Fourier dual of the Hartle-Hawking wave function (with real domains). However, even at the level of mini-superspace we notice an important
difference. If we impose that the torsion is strictly zero (i.e.  even off-shell and quantum mechanically), then we recover the Hartle-Hawking wave function. 
Allowing for off-shell torsion, however, changes the situation, a matter studied in detail in~\cite{QuantTorsion}. 

For simplicity let us set the spatial curvature to zero, ${k=0}$ (although it is not too hard to investigate the other cases). Then, the general ansatz for the connection consistent with 
homogeneity and isotropy is:
\bea
A^i_a&=&\delta^i_a(ib +c)\label{AMSS}\\
E^a_i&=&\delta^a_i a^2,\label{EMSS}
\eea
where $b$ and $c$ are functions of time. 
If the torsion is zero, then $b\approx \dot a$ and $c\approx 0$. Otherwise $b$ contains a parity-even component of the 
torsion, and $c$ a parity-odd component. The latter, the real part of $A^i$ in mini-superspace, is Cartan's spiral staircase~\cite{spiral,MZ}.
It must be zero as an equation of motion in Einstein-Cartan theory, but it may be switched on (classically) in quasi-topological theories of gravity~\cite{alex1,alex2,MZ}. 

With symmetry reduction (\ref{AMSS}) and (\ref{EMSS}) the Chern-Simons state (\ref{kodrealth}) reduces to:
\be\label{kod0}
\psi_{CS}={\cal N} \exp{\bigg(i\frac{ 3 V_c}{\Lambda l_P^2} \left(b^3 -3bc^2)\right)\bigg)}.
\ee
Reducing (\ref{EC}) to mini-superspace and inspecting the first (Legendre transform) term, leads to
\be
\big[\, \hat b\, ,\hat a^2\big]=i\frac{l_P^2}{3V_c}
\ee
where the extra factors of $3V_c$ result from a trivial integration over space and sum over indices $a$ and $i$.
Hence 
(\ref{FT}) reduces to:
\bea\label{FT1}
\psi_{a^2}(a^2)&=&\frac{\sqrt{3V_c}}{l_P}\int \frac{db }{\sqrt{2\pi}} e^{-i\frac{3V_c}{  l_P^2}a^2 b}\psi_b(b).
\eea
It is then a simple matter to show that~\cite{CSHHV}:
\be
\psi_{a^2}=  
{\cal N}'{\rm Ai}(-z),
\ee
with:
\be\label{zexp}
-z=\left(\frac{9V_c}{\Lambda l_P^2}\right)^{2/3}
\left(-c^2-\frac{\Lambda a^2}{3}\right).
\ee
Had we performed the calculation in a $k\neq 0$ model, the result would be the same but with:
\be\label{zexp}
-z=\left(\frac{9V_c}{\Lambda l_P^2}\right)^{2/3}
\left(k-c^2-\frac{\Lambda a^2}{3}\right).
\ee
As announced, if we force $c=0$, we recover the Hartle-Hawking wave function. However, if we do not force this off-shell, a different
picture emerges. The wave function seems to see an effective potential of the form:
\be\label{potential}
U(a)=4\left(\frac{3V_c}{l_P^2}\right)^2 a^2\left(k-c^2-\frac{\Lambda}{3}a^2\right),
\ee
that is, the usual one in the Wheeler-DeWitt equation, but with 
\be
k\rightarrow k-c^2.
\ee 
This is the crucial property that will allow us elsewhere to formulate a quantum version of the flatness problem and its possible 
solution~\cite{QuantFlat}. 

\section{Extension to anisotropic models}
We can now reduce the Chern-Simons state under whatever symmetry the problem has, and find the dual metric representation wave function.
We start  by illustrating this procedure with anisotropic models. 

\subsection{Bianchi I}
The procedure can be simply illustrated with the Bianchi I model for which the metric is:
\be
ds^2=-dt^2+a_i^2 (t) dx^2_i.
\ee
With standard formulae in the Ashtekar formalism~\cite{Thomas} this leads to:
\bea
E^a_i&=&\frac{1}{2}\delta^a_i \epsilon _{ijk}{\rm sgn} (a_i) a_j a_k\\
A^a_i&=&i \delta^a_i b_i
\eea
where on-shell (applying the torsion free condition) $b_i=\dot a_i$. The expansion rate in one direction is therefore conjugate to the geometrical average of the expansion rates in the {\it orthogonal} plane!  
We have:
\be
\{b_1,{\rm sgn} (a_1)a_2 a_3\}=\frac{1}{2\kappa V_c}
\ee
and cyclic perms, so that the corresponding quantum operators satisfy:
\be
\left[b_i,p_j\right]=i\delta_{ij}\frac{l_P^2}{ V_c}, \label{commutator}
\ee
with 
\bea
p_1 &=& \sign(a_1)a_2 a_3 \nonumber \\ 
p_2 &=& \sign(a_2)a_1 a_3 \nonumber \\ 
p_3 &=& \sign(a_3)a_1 a_2.
\eea
This is consistent with mini-superspace results, with a few interesting, but trivial points of note.
First, note that one drops a degeneracy factor (here the 3 equivalent directions in mini-superspace); this must happen every time one breaks a symmetry.
Second, the fact that the conjugate to the Hubble rate is $a^2$ and not $a$ (with well-known implications for the presence of a Euclidean branch)
appears to be an artefact of mini-superspace, since $a^2$ is replaced here by a product of expansion factors. 

The modified Chern-Simons state for this solution is:
\be\label{kod0}
\psi_{CS}={\cal N}  \exp{\bigg(i\frac{ 3 V_c}{\Lambda l_P^2} b_1 b_2 b_3\bigg)}
\ee
(where we will leave the ``normalization'' constant undefined for the time being)
and the Fourier transform implied by (\ref{FT}) is:
\bea\label{FT1}
\psi(a_1,a_2,a_3)&=&\left(\frac{V_c}{l_P^2}\right)^{3/2}\int \frac{db_1  db_2  db_3 }{(2\pi)^{3/2}} \times\nonumber\\
&& \times e^{-i\frac{V_c}{  l_P^2}(b_1 p_1 + b_2 p_2 +  b_3 p_3 )}\psi_{CS}.
\eea
Let us take the integrals over the whole real line, emulating the prescription for  Hartle-Hawking.

The integral can be solved as follows: one integral gives a delta function, the second integration is then trivial, so that
one is left with:
\be
\psi={\cal N} \frac{\Lambda }{3 l_P}\sqrt{\frac{V_c}{2\pi} } \int _{-\infty}^\infty \frac{dx}{|x|}e^{-i(x+C/x)}
\ee
with 
\be
C=\frac{\Lambda V_c^2}{3l_P^4}a_1^2 a_2^2 a_3^2{\rm sgn}(a_1 a_2 a_3)
\ee
(we will assume $\Lambda>0$ throughout).
This finally gives:
\be
\psi=
-{\cal N} \frac{\Lambda}{3l_P^2}\sqrt{2\pi V_c} Y_0\left(\frac{2 V_c}{l_P^2}
\sqrt{\frac{\Lambda}{3}}
a_1 a_2 a_3\right)
\ee
if $a_1 a_2 a_3 >0$.
If $a_1 a_2 a_3 <0$ the correct solution is:
\be
\psi=
{\cal N} \frac{4 \Lambda}{3l_P^2}\sqrt{\frac{V_c}{2\pi} }
K_0\left(\frac{2 V_c}{l_P^2}
\sqrt{\frac{\Lambda}{3}}
|a_1 a_2 a_3|\right)
\ee
where $Y_n(z)$ are the Bessel functions of the second kind and $K_n(z)$ are the modified Bessel functions of the second kind.

Note that the wave function is real, i.e. it is a stationary wave, just like the Hartle-Hawking.
However, it has a log divergence at the origin.
It is also a function of the factor that controls the volume (or in fact the volume squared). 

\begin{figure}
\begin{center}
  \includegraphics[width=10cm]{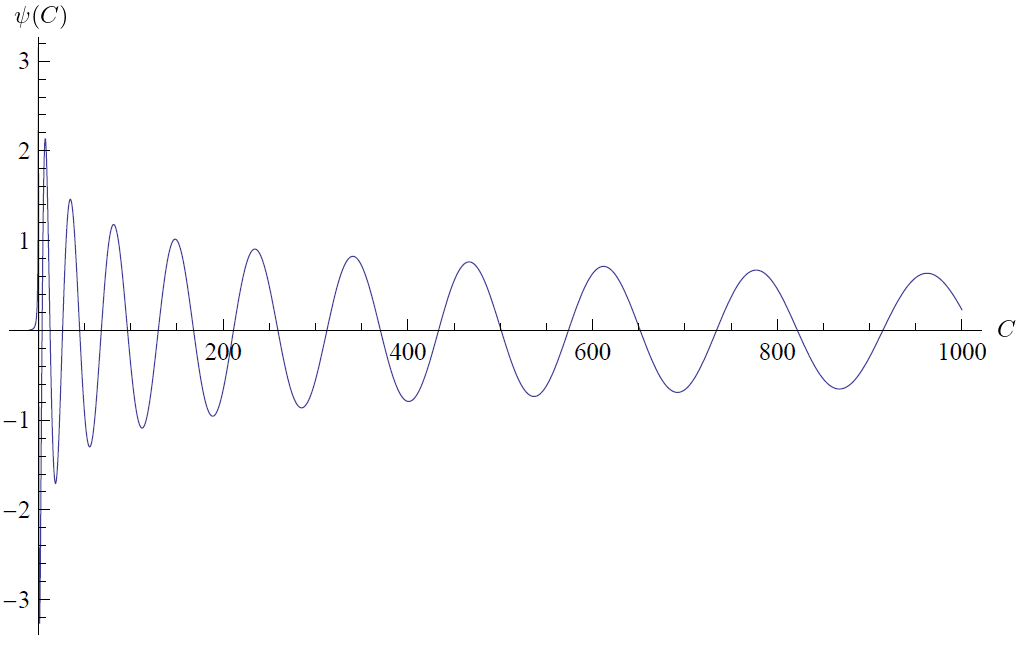}
  \caption{The ``Hartle-Hawking" wave function for the Bianchi I model, $\psi(C)$, for $C \in [-10, 1000]$.}
  \label{fig:longrun}
\end{center}
\end{figure}

\begin{figure}
\begin{center}
  \includegraphics[width=10cm]{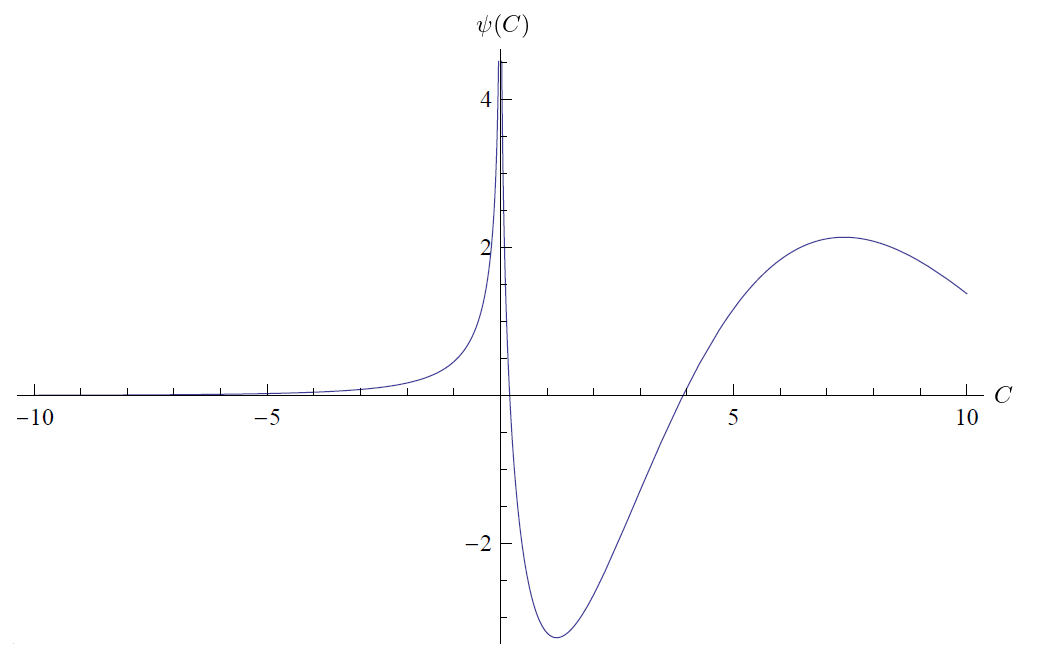}
  \caption{The ``Hartle-Hawking" wave function, $\psi(C)$ for $C \in [-10, 10]$.}
  \label{fig:zoom}
\end{center}
\end{figure}

\subsection{Bianchi I with quantum torsion}
In the previous examples, we have obtained a quantum state in the metric variables by imposing a torsion-free condition on the Ashtekar connection from the outset. However, it is worth emphasizing that in Ashtekar's formulation of general relativity, as in Cartan's formulation, the torsion-free condition on the connection is imposed as a second-class constraint, rather than as a kinematic restriction on the field space. While it was useful to impose the torsion-free condition in order to make contact with the Hartle-Hawking state (as well as the Vilenkin state), it may be propitious in other circumstances (e.g., within the Gupta-Bleuler formalism) to leave this constraint unsolved and allow for quantum torsion. Therefore, in this and the following sections, we consider a generalized version of the proposed ``Hartle-Hawking" state where we allow for a possibly non-vanishing quantum torsion. For Bianchi I, this amounts to including both real and imaginary parts for each of the connection components. While the imaginary parts are conjugate to the densitized triads, the real parts are simply parameters on which the Hartle-Hawking state will depend. 

We make the following ansatz for the connection:
\bea
A^1 &=& (ib_1 + c_1)dx \nonumber \\
A^2 &=& (ib_2 + c_2)dy \nonumber \\
A^3 &=& (ib_3 + c_3)dz, \label{connection1}
\eea
where $b_i$ and $c_i$ depend only on time. Evaluating $Y_{CS}$ on a hypersurface of constant time then gives
\bea
Y_{CS} &=& -\int A^1 A^2 A^3 \nonumber \\
\mathfrak{I}(Y_{CS})&=& -V_c(-b_1 b_2 b_3 + b_1 c_2 c_3 + b_2 c_1 c_3 + b_3 c_1 c_2) \nonumber \\
\eea
from which we get the modified Chern-Simons state
\be
\psi_{CS} = \mathcal{N}\exp\left\{-i\frac{3V_c}{\Lambda  l_P^2}\left(-b_1 b_2 b_3 + b_1 c_2 c_3 + b_2 c_1 c_3 + b_3 c_1 c_2\right)\right\}.
\ee
Taking the Fourier Transform gives the ``Hartle-Hawking" state

\bea
\psi(a_1,a_2,a_3)&=&\left(\frac{V_c}{l_P^2}\right)^{3/2}\int \frac{db_1  db_2  db_3 }{(2\pi)^{3/2}}\times \nonumber \\
&& \times  e^{-i\frac{V_c}{  l_P^2}(b_1 p_1 + b_2 p_2 +  b_3 p_3 )}\psi_{CS}. \nonumber \\
&=& \widetilde{\mathcal{N}}\int_{-\infty}^{\infty}\frac{dx}{|x|}e^{-i(x + C/x)} \nonumber \\
&=& \Bigg\{
    \begin{array}{cc}
    \!\!  -4\widetilde{\mathcal{N}}K_0(2\sqrt{|C|}), & C < 0 \\
     \!\! 2\pi\widetilde{\mathcal{N}} Y_0(2\sqrt{C}), & C > 0
    \end{array}
\eea
where
\begin{gather}
\widetilde{\mathcal{N}} = {\cal N} \frac{\Lambda }{3 l_P}\sqrt{\frac{V_c}{2\pi} }, \\
C = \frac{\Lambda}{3}\left(\frac{V_c}{ l_P^2}\right)^2(p_1 + \tfrac{3}{\Lambda}c_2 c_3)(p_2 + \tfrac{3}{\Lambda}c_1 c_3)(p_3 + \tfrac{3}{\Lambda}c_1 c_2).
\end{gather}
It is then clear that the net result of including quantum torsion is to shift the dependence of the wave function by ${p_i \to p_i + \tfrac{3}{\Lambda}c_j c_k}$, after which it is no longer a function of the spatial volume alone, but will also depend on the shape of space. In the context of homogenous, anisotropic models, it is often convenient to work with the so-called Misner variables $(\alpha, \beta_+, \beta_-)$:
\be
a_1 = e^{\alpha + \beta_+ + \sqrt{3}\beta_-}, \quad a_2 = e^{\alpha + \beta_+ - \sqrt{3}\beta_-}, \quad a_3 = e^{\alpha - 2\beta_+ }
\ee
so that $e^{3\alpha} = a_1 a_2 a_3$ characterizes the spatial volume density, and the anisotropy parameters $\beta_\pm$ describe the shape of space. In terms of the Misner variables, we can express the argument of the wave function as
\bea
C &=& \frac{\Lambda}{3}\left(\frac{V_c}{ l_P^2}\right)^2\bigg[e^{6\alpha} + \tfrac{3}{\Lambda}c_1 c_2 c_3 e^{4\alpha}\left(\tfrac{1}{c_1}e^{\beta_+ + \sqrt{3}\beta_-} + \tfrac{1}{c_2}e^{\beta_+ - \sqrt{3}\beta_-} + \tfrac{1}{c_3}e^{-2\beta_+} \right) \nn \\
&& + \left(\tfrac{3}{\Lambda}\right)^2 c_1 c_2 c_3 e^{2\alpha}\left(c_1 e^{-\beta_+ - \sqrt{3}\beta_-} + c_2 e^{-\beta_+ + \sqrt{3}\beta_-} + c_3e^{2\beta_+} \right) + \left(\tfrac{3}{\Lambda}\right)^3 c_1^2 c_2^2 c_3^2
\bigg].
\eea
Note that for real values of the Misner variables we must assume $a_i \geq 0$ so we have dropped the factors of $\sign(a_i)$ in the calculation above.
\subsection{Bianchi IX with quantum torsion}
Let us briefly review the basic features of the Bianchi IX model, focusing on the aspects that are most relevant to loop quantum cosmology (see, e.g. \cite{Corichi} and \cite{Battisti Marciano Rovelli}). The Bianchi IX model is a homogeneous but anisotropic generalization of \emph{closed} FRW cosmology. Each constant time hypersurface $\Sigma$ is assumed to have the topology of $\mathbb{S}^3$. The metric on $\Sigma$ is given by $d\Sigma^2 = \sum\limits_i a_i(t)^2 (\omega^i)^2,$ where
\bea
\omega^1 &=& \cos\psi d\theta + \sin\psi\sin\theta d\phi \nonumber \\
\omega^2 &=& \sin\psi d\theta - \cos\psi\sin\theta d\phi \nonumber \\
\omega^3 &=& d\psi + \cos\theta d\phi
\eea
and where $\psi \in (0, 4\pi)$, $\phi \in (0, 2\pi)$, and $\theta \in (0, \pi)$ are coordinates on $\mathbb{S}^3$. It is convenient to note that $\omega^i$ are Maurer-Cartan forms satisfying the relations $d\omega^I = \tfrac{1}{2}\epsilon^I{}_{\!JK}\omega^J \omega^K$.

The Ashtekar connection is given by 
\bea
A^1 &=& (ib_1 + r_1)\,\omega^1 \nonumber \\
A^2 &=& (ib_2 + r_2)\,\omega^2 \nonumber \\
A^3 &=& (ib_3 + r_3)\,\omega^3 \label{connection9}
\eea
where $r_i = -\Gamma_i + c_i$ are the real parts of the connection, including both the torsion-free component $\Gamma_i$ and the torsion $c_i$. We can make a convenient choice of physical frame
\be
e^1 = a_1\omega^1, \quad
e^2 = a_2\omega^2,  \quad
e^3 = a_3\omega^3
\ee
from which we can infer the torsion-free part of the spatial connection
\bea
\Gamma_1 &=& \frac{1}{2}\left(\frac{a_2}{a_3} + \frac{a_3}{a_2} - \frac{a_1^2}{a_2 a_3}\right) \nonumber \\
\Gamma_2 &=& \frac{1}{2}\left(\frac{a_1}{a_3} + \frac{a_3}{a_1} - \frac{a_2^2}{a_1 a_3}\right) \nonumber \\
\Gamma_3 &=& \frac{1}{2}\left(\frac{a_1}{a_2} + \frac{a_2}{a_1} - \frac{a_3^2}{a_1 a_2}\right)
\eea
and the densitized triad
\bea
E_1 &=& \frac{p_1}{16\pi^2}\left( -\sin\psi\cos\theta \partial_\psi + \cos\psi\sin\theta\partial_\theta + \sin\psi\partial_\phi \right) \nonumber \\
E_2 &=& \frac{p_2}{16\pi^2}\left( \cos\psi\cos\theta \partial_\psi + \sin\psi\sin\theta\partial_\theta - \cos\psi\partial_\phi \right) \nonumber \\
E_3 &=& \frac{p_3}{16\pi^2} \partial_\psi, \label{triads9}
\eea
with $p_i$ defined as in Bianchi I. Now, the Poisson bracket relations become
\be
\{b^i, p_j\} = \frac{1}{2\kappa} \delta^i_{\,j},
\ee
which, upon quantization, lead to the commutations relations
\be
[\hat{b}^i , \hat{p}_j] = i  l_P^2 \delta^i_{\,j}.
\ee
We can also express the torsion-free part of the connection in terms of the momenta $p_i$ via
\bea
\Gamma_1 &=& \frac{1}{2}\left(\frac{p_2}{p_3} + \frac{p_3}{p_2} - \frac{p_2p_3}{p_1^2}\right) \nonumber \\
\Gamma_2 &=& \frac{1}{2}\left(\frac{p_1}{p_3} + \frac{p_3}{p_1} - \frac{p_1p_3}{p_2^2}\right) \nonumber \\
\Gamma_3 &=& \frac{1}{2}\left(\frac{p_1}{p_2} + \frac{p_2}{p_1} - \frac{p_1p_2}{p_3^2}\right).
\eea
Computing the Chern-Simons functional now proceeds along the same lines as in Bianchi I, except that now the term proportional to $A^I dA^I$ no longer vanishes. 
\bea
Y_{CS} &=& -\int\tfrac{1}{2}A^I dA^I + A^1 A^2 A^3 \\
\mathfrak{I}(Y_{CS})&=& -16\pi^2[(b_1r_1 + b_2r_2 + b_3r_3) + \\
&& + (-b_1 b_2 b_3 + b_1 r_2 r_3 + b_2 r_1 r_3 + b_3 r_1 r_2)]. \nonumber
\eea
We can again form the modified Chern-Simons state
\be
\psi_{CS} = \mathcal{N}\exp\left\{-i\frac{48\pi^2}{\Lambda  l_P^2}\left[(b_1r_1 + b_2r_2 + b_3r_3) + (-b_1 b_2 b_3 + b_1 r_2 r_3 + b_2 r_1 r_3 + b_3 r_1 r_2)\right]\right\}.
\ee
Taking the Fourier transform gives the ``Hartle-Hawking" state, which leads once again to the same integral
\be
~~~\hspace{1cm} \nonumber 
\ee
\bea
\psi(a_1 ,a_2, a_3) &=& \frac{\mathcal{N}}{l_P^3}\int \frac{db_1  db_2  db_3 }{(2\pi)^{3/2}} \times \nonumber \\
&& \times  e^{-\frac{i}{l_P^2}(b_1 p_1 + b_2 p_2 +  b_3 p_3 )}\psi_{CS} \nonumber \\
&=& \widetilde{\mathcal{N}}\int_{-\infty}^{\infty}\frac{dx}{|x|}e^{-i(x + C/x)} \nonumber \\
&=& \Bigg\{
    \begin{array}{cc}
    \!\!  -4\widetilde{\mathcal{N}}K_0(2\sqrt{|C|}), & C < 0 \\
     \!\! 2\pi\widetilde{\mathcal{N}} Y_0(2\sqrt{C}), & C > 0
    \end{array}
\eea
where we now have:
\begin{gather}
\widetilde{\mathcal{N}} = \frac{\mathcal{N}\Lambda}{48\pi^2\sqrt{2\pi}l_P}, \\ 
C = \frac{\Lambda l_P^2}{48\pi^2}\left[\frac{p_1}{l_P^2} + \frac{48\pi^2}{\Lambda l_P^2}(r_1 + r_2 r_3)\right]\left[\frac{p_2}{l_P^2} + \frac{48\pi^2}{\Lambda l_P^2}(r_2 + r_1 r_3)\right]\left[\frac{p_3}{l_P^2} + \frac{48\pi^2}{\Lambda l_P^2}(r_3 + r_1 r_2)\right].
\end{gather}
We can define Misner variables for Bianchi IX analogous to those used in Bianchi I accounting for the fact that the scale factors, which were dimensionless in Bianchi I, now have units of length:
\be
\frac{a_1}{l_P} = e^{\alpha + \beta_+ + \sqrt{3}\beta_-}, \quad \frac{a_2}{l_P} = e^{\alpha + \beta_+ - \sqrt{3}\beta_-}, \quad \frac{a_3}{l_P} = e^{\alpha - 2\beta_+ }.
\ee
In terms of these Misner variables, the argument of the Hartle-Hawking wave function becomes
\be
C = \frac{\Lambda l_P^2}{48\pi^2}\left[\frac{e^{2\alpha - \beta_+ - \sqrt{3}\beta_-}}{8\pi} + \frac{48\pi^2}{\Lambda l_P^2}(r_1 + r_2 r_3)\right]\left[\frac{e^{2\alpha - \beta_+ + \sqrt{3}\beta_-}}{8\pi} + \frac{48\pi^2}{\Lambda l_P^2}(r_2 + r_1 r_3)\right]\left[\frac{e^{2\alpha + 2\beta_+ }}{8\pi} + \frac{48\pi^2}{\Lambda l_P^2}(r_3 + r_1 r_2)\right].
\ee

\section{The Kantowski-Sachs model}
Finally, we consider the generalized Hartle-Hawking state for the Kantowski-Sachs model \cite{Kant}, which may be viewed as a homogeneous, anisotropic cosmological model, or as a reduced phase space for the interior of a spherically symmetric black hole. Indeed, the Kantowski Sachs model generalizes the interior of a Schwarzschild black hole, which can be seen by inspecting the line element
\be
ds^2 = - dt^2 + a_1(t)^2 dx^2 + a_2(t)^2(d\theta^2 + \sin^2\theta d\phi^2). \label{Kantowski-Sachs}
\ee
In order to make the connection with Schwarzschild manifest, one can use the function $r = a_2(t)$ as a ``radial" coordinate, in which case the line element becomes
\be
ds^2 =    -A(r)^2 dr^2 +  B(r)^2 dx^2 + r^2(d\theta^2 + \sin^2\theta d\phi^2), \label{Kantowski-Sachs-radial}
\ee
for some functions $A(r)$ and $B(r)$. If one chooses $A(r)^{-2} = B(r)^2 = \tfrac{2m}{r}-1$, this becomes the interior of the Schwarzshild spacetime, where the coordinate $x$ is identified as the time coordinate in the exterior portion of the spacetime. Now let us return to the line element \eqref{Kantowski-Sachs} in the original coordinate system. The hypersurfaces of constant $t$ have topology $\mathbb{S}^2 \! \times \! \mathbb{R}$, with $(\theta,\phi)$ being standard coordinates on $\mathbb{S}^2$, and $x$ a coordinate on $\mathbb{R}$. Since the spatial sections are non-compact, we introduce the fiducial length scale $ l_0$ and restrict $x \in (0, l_0)$.  Following \cite{Joe and Singh} we make a convenient choice of physical frame
\be
e^1 = -a_2\sin\theta d\phi, \quad e^2 = a_2 d\theta, \quad e^3 = a_1 dx,
\ee
from which we can find the torsion-free connection
\be
\omega^1{}_2 = -\cos\theta d\phi, \quad \omega^1{}_3 = \omega^2{}_3 = 0
\ee
and the densitized triad
\bea
E_1 &=& -p_1 \partial_\phi \nonumber \\ 
E_2 &=& p_1\sin\theta\partial_\theta \nonumber \\ 
E_3 &=& p_2\sin\theta\partial_x,
\eea
where $p_1 = a_1 a_2$ and $p_2 = a_2^2$. For the Ashtekar connection, we make the corresponding ansatz
\bea
A^1 &=& -(ib_1 + c_1)\sin\theta d\phi \nonumber \\ 
A^2 &=& (ib_1 + c_1)d\theta \nonumber \\ 
A^3 &=& (ib_2 + c_2)dx + \cos\theta d\phi. \label{A-Kantowski-Sachs}
\eea
where the $\cos\theta d\phi$ term in $A^3$ is the only torsion-free part of the connection and $c_1$, $c_2$ are torsion components. From the triad and connection, we deduce the Poisson bracket relations
\be
\{b_1, p_1\} = \frac{1}{16\pi \kappa l_0}, \quad \{b_2, p_2\} = \frac{1}{8\pi\kappa l_0},
\ee
where once again, $c_1$, $c_2$ Poisson-commute with all phase space variables. Quantization yields the following commutation relations:
\be
[\hat{b}_1, \hat{p}_1] = i\frac{ l_P^2}{8\pi l_0}, \quad [\hat{b}_2 , \hat{p}_2] = i\frac{l_P^2}{4\pi l_0}.
\ee
The Chern-Simons functional for the connection \eqref{A-Kantowski-Sachs} becomes
\bea
Y_{CS} &=& -\int \frac{1}{2} A^3 dA^3 + A^1 A^2 A^3  \nonumber \\
&=& 4\pi l_0\left(\frac{i}{2}b_2 + i b_2 b_1^2 + 2 b_1 b_2 c_1 - i b_2 b_1^2 + \frac{1}{2} c_2 + c_2 b_1^2 - 2ib_1 c_1 c_2 - c_1 ^2 c_2 \right),
\eea
and taking the imaginary part, we have
\be
\mathfrak{I}(Y_{CS}) = 4\pi l_0\left( \frac{1}{2}b_2 + b_2 b_1^2 - b_2 c_1^2 - 2 b_2 c_1 c_2 \right).
\ee
We can now form the modified Chern-Simons state for the Kantowski-Sachs model:
\bea
\psi_{CS} &=& \mathcal{N}\exp\left( \frac{3i}{ l_P^2 \Lambda}\mathfrak{I}(Y_{CS}) \right) \nonumber  \\
&=&  \mathcal{N}\exp\left\{ \frac{3i}{ l_P^2 \Lambda}  4\pi l_0\left( \frac{1}{2}b_2 + b_2 b_1^2 - b_2 c_1^2 - 2 b_2 c_1 c_2  \right)  \right\}.
\eea
Taking the Fourier transform leads to the Hartle-Hawking state
\bea
\psi(a_1, a_2) &=& \sqrt{2}\mathcal{N}\frac{4\pi l_0}{l_P^2}\int\frac{db_1 db_2}{2\pi}\exp\left\{-i\frac{ 4\pi l_0}{l_P^2}\left(2b_1 p_1 + b_2 p_2 \right) + \frac{3i}{ l_P^2\Lambda}4\pi l_0\left( \frac{1}{2}b_2 + b_2 b_1^2 - b_2 c_1^2 - 2 b_1 c_1 c_2 \right)\right\} \nonumber \\
&=& \frac{\widetilde{\mathcal{N}}}{\sqrt{c_1^2 - \frac{1}{2} + \frac{\Lambda}{3}p_2}}\exp\left\{-i\frac{8\pi l_0}{ l_P^2}\left(p_1 + \frac{3}{\Lambda}c_1 c_2\right)\sqrt{c_1^2 - \frac{1}{2} + \frac{\Lambda}{3}p_2}\right\}, \qquad  \widetilde{\mathcal{N}} = \frac{\mathcal{N}\Lambda }{3\sqrt{2}}.
\eea

\section{Outlook}

The importance of the tool presented in this paper is twofold. Firstly, it allows us to generate solutions to the Wheeler-DeWitt equation in the metric representation not by solving a differential equation, but by computing a Fourier transform. By realizing that the Chern-Simons state is the solution in the connection representation for spaces of any symmetry, and that the metric representation is just the Fourier transform of the connection representation, we find a shortcut for generating solutions that generalize the Hartle-Hawking state for any situation. We presented examples related to Bianchi and Kantowski-Sachs models, but the applications are endless.  Inhomogeneous cosmological models, black holes, and exact gravitational waves spring to mind. In all cases, we have already provided a formal solution: and that is  (\ref{FT}) with (\ref{kodrealth}).  All that remains to be done is to simplify and interpret this solution on a case by case basis. In principle, the methods illustrated in this article should be applicable to more realistic settings for the early universe wavefunction, including, for example, the case of a gauge field supported wavefunction for a radiation dominated universe, as studied in \cite{Bertolami:1991cf}.

Secondly, by starting from the connection representation, we open up the doors to torsion. Even if this is eventually found to be zero
classically (or ``on-shell''), the quantum theory should be able to probe torsion off-shell. However, torsion degrees of freedom are frozen by construction in the metric formulation. The fact that we can insert torsion degrees of freedom into ansatze of any symmetry allowed us to find wave functions in the metric representation which take torsion into account.  It is intriguing that the torsion correction to the wave function modifies the effective potential to correct the curvature, opening a window for addressing the curvature problem at the quantum level.  We leave this prospect for future investigation~\cite{QuantFlat}.  Another avenue concerns singularity avoidance.  A close inspection of the commutation relations for the Bianchi I model, eq \eqref{commutator}, reveals that the expansion rate in one principal direction is conjugate to the geometrical average of the expansion rates in the orthogonal plane.  In other words, in a manner reminiscent of any simple quantum mechanical system, e.g. the harmonic oscillator, where the quantum mechanical uncertainty in position cannot approach zero unless the uncertainty in the momentum diverges, the uncertainty of the scale factor will evade going to zero unless the curvature diverges. It may be that the canonical commutation relations between the scale factors and the connection variables are precisely what is needed to ensure singularity avoidance at the big bang. In fact, singularity avoidance in the context of the Bianchi I model has already been studied in \cite{Kiefer Kwidzinski Piontek}, but the relationship between the modified CS state and the generalized Hartle-Hawking state we have explored here may provide some additional insights in this direction. We leave a more detailed analysis of singularity avoidance for future work.  
 
Finally, the Chern-Simons state may also represent the Vilenkin wave function, depending on the choice of contour, should we allow excursions into the complex domain for variables usually taken to be real~\cite{CSHHV}. The fact that boundary conditions in one representation translate into contours in the other is reminiscent of the discussions leading to the Feynman propagator. It would be very interesting to investigate this matter further, but note that issues of convergence of the Fourier integral necessarily come into play, making this enterprise non-trivial.

\section*{Acknowledgements}

We would like to thank Lee Smolin for discussions, and Steve Carlip for helpful feedback.  
This work was supported by the STFC Consolidated Grant ST/L00044X/1 (JM).

\end{document}